# C.V. Raman's Exploration in Optics: A Spectrum of History

G.V. Pavan Kumar

Department of Physics, Indian Institute of Science Education and Research, Pune – 411008, India

pavan@iiserpune.ac.in

## Table of Contents

# Abstract

C.V. Raman (1888-1970) was one of the pioneering scientists to have emerged from India during the colonial era. His scientific explorations were driven by his curiosity to understand wave phenomena. Naturally, optics and related physical effects were at the heart of such an exploration. Apart from his Nobel prize-winning discovery of the Raman effect, his research included topics such as oblique diffraction, light scattering from liquids and amorphous solids, classical and quantum nature of light, acousto-optics, haloes and coronae (speckles), crystal dynamics and soft modes, optics of minerals, floral colors, physiology of vision and many other aspects related to light in natural settings. In this article, I give a historical overview of some of the work by C.V. Raman and his group that had a direct connection to optics and optical spectroscopy.

## Introduction

Human minds are driven by curiosity. It motivates them to explore natural phenomena and ask questions. If these questions are systematically pursued, then it can lead to the discovery of new scientific facts and ideas. Chandrasekhar Venkat Raman (C.V. Raman) had an abundance of such curiosity, which drove him to become one of the most recognized scientists from India. At the heart of Raman's curiosity-driven research were natural phenomena connected to optics and the related physics of waves. During his career, which spanned over 65 years, Raman wrote more than 400 research papers [1–6]. His writings also include monographs, technical lectures and public science talks [7]. Throughout his life, Raman engaged in a rich scientific life with a spirit of inquiry and an enthusiasm to discover new scientific phenomena.

In the pursuit of scientific discovery, one will have to consider the facts and ideas. They contribute to each other's development, and generally, the result is greater than the sum of their individual parts. Raman's motivation was to discover new scientific phenomena aided by facts and ideas. He had a deep appreciation of experimental and theoretical contributions towards this endeavour. He thought deeply about the process of scientific research, and he succinctly describes it in a lecture as follows [8]: "What is meant by a scientific discovery? How is it made? These are questions of perennial interest which are often asked and to which the most varied answers have been returned. A discovery may obviously be either of a new fact or of a new idea. It is clear however that an unexplained observation is of no particular significance to science. An idea unsubstantiated by facts is equally devoid of importance. Hence to possess real significance a scientific discovery must have both an experimental and a theoretical basis. Which of these aspects is the more important depends on the particular circumstances of the case, and a rough distinction thereby becomes possible between experimental and theoretical discoveries."

The aim of this paper is to give a glimpse of Raman's discovery, perhaps a metaphorical spectrum of history, related to optical phenomena. The paper is not an exhaustive summary of Raman's work, but an overview of certain ideas that I found interesting and may have contemporary connections to optics and photonics. I give particular emphasis on how ideas in optics influenced Raman's work throughout his career, including the discovery of the effect named after him.

## Timelines

### Raman's life

Chandrasekhara Venkata Raman (C.V. Raman) was born on 7th November 1888 in Thiruvanaikkaval in Tamil Nadu (a southern state of India), colonial India. His mother was Parvathi Ammal, and his father was R. Chandrasekhara Iyer. The couple had eight children, and Raman was the second child. Mr. Iyer was a local school teacher and later became a lecturer in mathematics and physics. After primary education, Raman, at the age of 13, joined the Presidency College in Madras. At the age of 15, he passed the BA degree with first class. He obtained a gold medal in Physics and English. It was also during his undergraduate years that Raman was exposed to independent research, and the work done in Presidency College resulted in a couple of research papers (more on this in the next section). By 1906, he had completed his MA degree and was contemplating traveling abroad for higher education. Upon medical examination by the civil surgeon in Madras, he was discouraged from traveling abroad, especially to colder environments, and was advised not to risk his health. In 1907, he qualified for the Indian Financial Service and was posted as Assistant Accountant General in Calcutta (eastern part of India). Before moving to Calcutta, Raman got married to Lokasundari Ammal. From 1907 to 1933, Raman spent most of his time at Calcutta (with brief periods at Rangoon and Nagpur). In the initial decade, he was still employed in the Indian financial service and performed independent research at the Indian Association for the Cultivation of Science (IACS). In 1917, he quit the Indian Financial Service and became a professor at Calcutta University and IACS. It is during this era that Raman and his students performed the groundbreaking experiments on Raman scattering. He went on to win a Nobel Prize in Physics in 1930. In 1933, Raman was appointed as Director (the first Indian) of the Indian Institute of Science (IISc) at Bangalore. In mid-1937, he quit the directorship of IISc and continued as a professor at the same place till 1948. From 1948 to 1970, Raman was the Director of the Raman Research Institute at Bangalore. Raman passed away on 21st November 1970 in Bangalore. For the interested readers, I suggest two (scientific) biographies of C. V. Raman: one is written by Venkataraman [9], and the other (with a lot of pictures) is by Ramaseshan and Rao [10].

## Raman's Career

Raman's research career can be divided into four phases.

1. 1907-1917  Indian Financial Service & An independent scientist at the Indian Association for the Cultivation of Science (IACS), Calcutta
2. 1917 – 1933 Palit Professor (Calcutta University and IACS)
3. 1933-1937  Director & Professor (Indian Institute of Science, Bangalore)
4. 1937-1948  Professor (Indian Institute of Science, Bangalore)
5. 1948-1970 Director & Professor (Raman Research Institute, Bangalore)

## Raman's research by topic

As an undergraduate student, Raman's research interests started in optics and evolved over his lifetime. The common theme was to explore the wave phenomena in light and sound. As noted by Ramaseshan and Rao (see p. 174 in reference [10]), the timeline of Raman's research interests can be roughly divided as follows:

1. Acoustics (1910-1920)
2. Optics and scattering of light (1920-1930),
3. Ultrasonic diffraction and the application of Brillouin scattering to liquids and Raman scattering to crystals (1930-1940)
4. Diamond and vibrations of crystal lattices (1940-1950)
5. Optics of minerals (1950-1960)
6. Colours and their Perceptions (1960-1970)

With the above timeline as a reference, the motivation of this article is to give a glimpse of Raman's work on optics.

## Explorations in Optics

Understanding the interaction between light and matter was one of the main interests of C.V. Raman. Naturally, optical phenomena intrigued him. He studied a variety of problems in classical optics, including phenomena such as interference fringes, oblique angle diffraction, haloes, coronae, speckles, aberration, wave fronts and many more (see collection of papers in [3]). It is not possible to cover all these topics within the scope of this article, but I will highlight some of the papers that I found interesting and have implications in contemporary optics and photonics.

## Formative Studies (1906 -1920)

Ever since his undergraduate years, Raman had been interested in vibrations and waves. This influenced him to study acoustics and optics in elaborate detail. The wave phenomena were central to most of his work. He was deeply influenced by theoretical and experimental developments in optics. He studied the work of Newton, Young, Fresnel, Fraunhofer, Huygens, and Grimaldi (see early chapters of [9] and [10]). Specifically, he was well acquainted with Huygens' wavelet theory and extensively utilized the work of Kirchhoff and Sommerfeld in his work related to optics.

Raman was not only exposed to and explored the theoretical developments of his era, but he was also interested in experimental optics, including instrumentation. So much so that he carried a Nicol prism in his pocket during his journey across the seas and utilized it to perform observations on board. Throughout his career, Raman was interested in understanding how waves behave under a variety of conditions, and therefore, optics and acoustics were natural platforms to study them. It is not surprising that he eventually worked on acousto-optics, where the wave phenomena from light and sound came into combined reasoning. That part of his contribution will be discussed later.

The very first paper by C. V. Raman was written in the year 1906, and the title of the paper was "*Unsymmetrical diffraction bands due to a rectangular aperture*", published in Philosophical Magazine [11]. C. V. Raman was an undergraduate student (~18 years), and his affiliation in the paper reads as "demonstrator in physics, Presidency College, Madras." This paper experimentally investigated the diffraction pattern of light obliquely illuminated on rectangular apertures. Specifically, he was interested in the diffraction bands that were rendered asymmetric due to oblique incidence. The paper presents a systematic experimental study by using two different experimental configurations (slit and prism). One of the intriguing observations that Raman made was that the nature of the intensity distribution changes as the oblique incidence angle changed from 85° to 87°. Of course, he did not give a precise explanation for this, but put forth an observation that needed further analysis. In the future course of his career, he came back to this problem to address related questions in the asymmetric band distribution due to the oblique incidence of light on apertures. It is evident from the paper that he was knowledgeable in classical optics, including the wave theory of light, and was comfortable in communicating ideas in a systematic way. It also showcases the experimental capability of C.V. Raman, which played a critical role throughout his career. We need to remember that Raman was just an 18-year-old student, and writing an experimental optics paper at that age is a significant achievement. It was a study that indicated his intellectual independence, a hallmark throughout his career.

Raman started as an independent, part-time researcher working in Calcutta. His main employment was in Indian financial services. Gradually, a research group was built around him. A band of enthusiastic students joined him to perform experimental research. A majority of them worked on problems related to wave phenomena, particularly in optics. During the late 1910s and early 1920s, Raman worked on acoustical effects in musical instruments [2] and gradually shifted towards research on optics [3,12]. A theme he revisited during this era was oblique diffraction, which was the theme of his very first research paper. At the Indian Association of Cultivation of Science, he introduced photographic techniques into his experimental repertoire, and this played an important role in observing a variety of optical phenomena. Problems, including diffraction fringes from oblique incidence experiments [12], were photographed with reasonable clarity and accuracy, and this led to further development of the associated theory.

## Raman's epochal decade (1920 – 1930)

### Discovery of the Raman Scattering

The origin of Raman's research on light scattering is connected to his observations made on board a ship while travelling to Europe [13]. He was fascinated to understand the blue colour of the Mediterranean Sea and was motivated to test the observations in a laboratory setting. This initial curiosity-driven research took him to paths that were unanticipated. He gradually realized that a systematic study of light scattering had a broad scope. It tested not only foundational concepts of an optical phenomenon but also connected to deeper problems in physics and chemistry. As he describes in his Nobel lecture [13]:

"A voyage to Europe in the summer of 1921 gave me the first opportunity of observing the wonderful blue opalescence of the Mediterranean Sea. It seemed not unlikely that the phenomenon owed its origin to the scattering of sunlight by the molecules of the water. To test this explanation, it appeared desirable to ascertain the laws governing the diffusion of light in liquids, and experiments with this object were started immediately on my return to Calcutta in September 1921. It soon became evident, however, that the subject possessed a significance extending far beyond the special purpose for which the work was undertaken, and that it offered unlimited scope for research. It seemed indeed that the study of light-scattering might carry one into the deepest problems of physics and chemistry, and it was this belief which led to the subject becoming the main theme of our activities at Calcutta from that time onwards."

The first quarter of the 20th century was an era when quantum mechanics emerged as an intriguing aspect of physics [14]. One of the fundamental observations related to quantum phenomena, especially with respect to light-quantum matter interaction, was observed by Arthur Compton [15]. The discovery of Raman scattering was motivated by the observation of the Compton effect. In the 1920s, Arthur Compton experimentally studied the scattering of X-rays from free electrons. He observed that in addition to the elastic component of the scattered X-rays, there was a detectable amount of inelastic component in the X-ray spectrum. This inelastic component was generally termed the modified component of the scattered radiation. This effect is called the Compton effect, and in those years, it led to an important insight into the interaction of light with quantum matter, particularly with free electrons. Until then, the experimental evidence for such interactions was few, and the theoretical developments were in the nascent stage. Therefore, Compton's observation came as an important clue to understand light-quantum matter interaction.

In Compton's experiment, the inelastic component of the scattered radiation played a crucial role in understanding the exchange of energy processes between the input light and the electron. Raman was aware of these developments, and it motivated him to look for similar evidence in the optical band of electromagnetic waves.

This is evident from the opening lines of the famous paper [16] titled "A new type of secondary radiation" by C. V. Raman and K. S. Krishnan, which announced the inelastic scattering process for visible light:

"If we assume that the X-ray scattering of the unmodified type observed by Professor Compton corresponds to the normal or average state of atoms and molecules, while the modified scattering of the altered wavelength corresponds to their fluctuation from that state, it would follow that we should expect also in the case of ordinary light, two types of scattering: one determined by the normal optical properties of the atoms or molecules, and another representing the effect of their fluctuation from the normal state. "

There are two aspects that are noteworthy: Raman and Krishnan draw an analogy between X-ray scattering and optical scattering. The second aspect is that the inelastic scattering component, which they term the 'modified scattering', is connected to the fluctuations from the normal state of atoms and molecules. Of course, they do not go into the details of what exactly these fluctuations mean, but it is an important clue to get an insight into the dynamics of atomic and molecular systems. This aspect is one of the most important observations in the context of *visible light*-quantum matter interaction.

Given that the inelastic scattering component of light was weak, Raman and his group members had to ensure a high intensity of light scattering. Note that this was the pre-laser era. The problem of intensity was tackled by focusing sunlight onto the scattering sample. To achieve illumination conditions close to monochromaticity, colour filters were carefully chosen and arranged for every experiment. The scattering samples (liquids and vapours) had to be purified and ensured to be dust-free, which was a technical challenge. By using complementary filters, only the modified scattering was first detected through visual observation and subsequently confirmed by spectroscopic measurements. The initial report states [9] that about 60 liquid samples were test to ensure the scattering phenomenon.

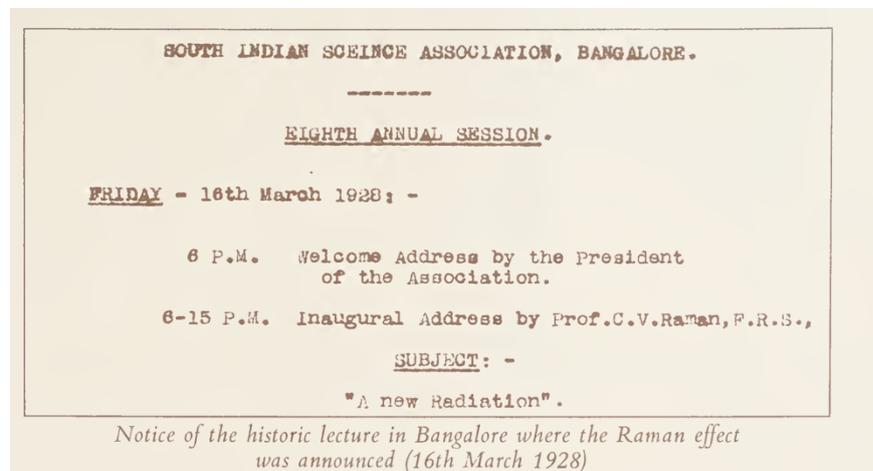

Figure 1. "Notice of the historic lecture in Bangalore where the Raman effect was announced". Image and caption reproduced from reference [10] under Creative Commons license.

The phenomenon of fluorescence was already known at this time. It was also common knowledge that conventional fluorescence resulted in a decrease in the frequency (or an increase in the wavelength).It was important for Raman and his group to confirm that the modified component that they were measuring was indeed coming out of a scattering phenomenon and not from fluorescence. Thus, Raman and Krishnan had to differentiate the scattered light from the fluorescence signal. This was done by determining the polarization of the scattered light, and the paper reports thus [9]:

"That the effect is a true scattering and not a fluorescence is indicated in the first place by its feebleness in comparison with the ordinary scattering, and secondly by its polarisation, which is in many ways quite strong and comparable with the polarisation of the ordinary scattering."

The year 1928 is an important one in the context of the discovery of the Raman Effect. Raman's group was actively pursuing this research area, and there were important papers that elaborated further on this topic [17–21,13]. Among them, one of the papers that gave details of the observation of the so-called modified scattering was published in the *Indian Journal of Physics* [22]. This was a reprint of the lecture that C V Raman gave in Bangalore (see figure 1) and is titled "*A New Radiation.*" (for specific details on dates, see [9])

Raman introduces the concept of the normal and secondary radiation and explains why the observed effect is a new phenomenon. He discusses the universality of this observation, emphasizing the physical mechanism behind the light-matter interaction. Of particular interest and importance is the line spectrum of the new radiation, which Raman and his research group had been recording and studying over the past few years. This paper is also one of the first locations in which the Raman spectra were shown (see figure 2) with photographic plates [22].

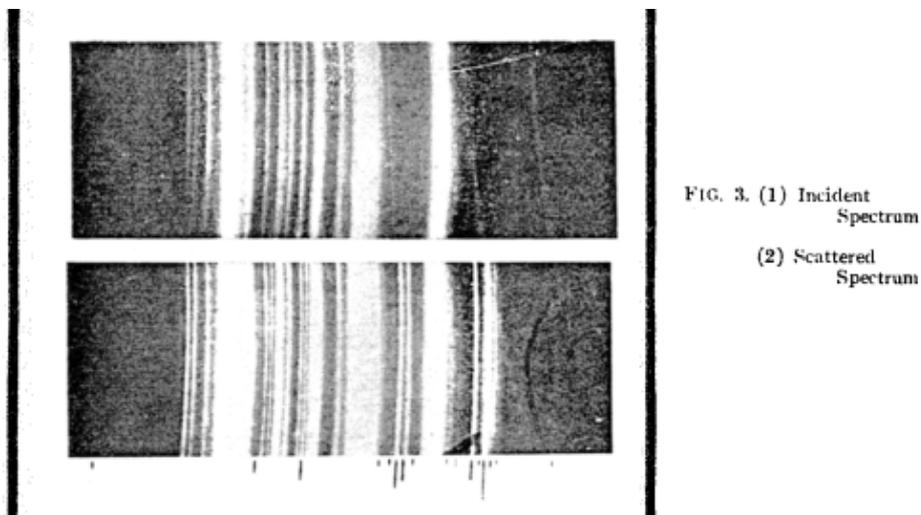

Figure 2. One of the first Raman scattering spectra ever published. Reproduced from reference [22] under Creative Commons License. Raman and Krishnan describe the figures as follows: "Figs. 3 (1) and 3 (2)...show the phenomenon. They are spectrograms taken with a small Hilger quartz instrument of the scattering by liquid benzene. Fig. 3 was taken with the light from the quartz mercury arc filtered through a blue glass which allows the wavelengths from about 3,500 A.U. to 4,400 A.U. to pass through. Fig. 3 (1) represents the incident spectrum and Fig. 3 (2) the scattered

spectrum, and the latter shows a number of sharp lines not present in Fig. 3 (1). These are indicated in the figure."

Raman also discusses the nature of the new radiation and its connection to thermodynamics. He asks what might be the nature of the radiation that is scattered from this interaction? And if the emerging waves are coherent or otherwise? Furthermore, he draws an analogy to the x-ray scattering that Compton had observed and finally concludes the paper with the following remark: "We are obviously only at the fringe of a fascinating new region of experimental research which promises to throw light on diverse problems relating to radiation and wave-theory, X-ray optics, atomic and molecular spectra, fluorescence and scattering, thermodynamics and chemistry. It all remains to be worked out." (see p. 376 in [22])

This anticipation has indeed turned out to be correct because Raman scattering is a very powerful tool to study various states of matter. Its application in chemistry and biology has been significant, so much so that it has been employed in medical diagnostics.

For those interested, an elaborate history of the discovery can be found in many sources [9,10,23–25]. I will also encourage readers to have a look at the Diary entry of K. S. Krishnan [26], which gives a day-by-day account of the discovery. The 28th of February is celebrated as National Science Day in India in commemoration of the discovery of the Raman Effect.

I would also like to point out an important contribution towards inelastic scattering of light in solids by Mandelstam, which was contemporary to Raman's discovery. Raman's observations were mainly in liquids, and his publication predates that of Mandelstam by a few weeks. A good account of the Russian contribution and the related mini-controversy with respect to the prominence of the discovery can be found in Raman's scientific biography (see chapter 5 in [9]), and the Russian perspective can be found in the paper by Fabelinski [27].

The nature of light, especially the photon picture, was still in its infancy during the 1920s and 30s. Raman and his research group (particularly Bhagavatam) took concrete efforts [28–30] to gain greater clarity on energy, momentum and spin of photons in particular. This was an attempt to extrapolate an electromagnetic picture of Maxwell's field theory from the classical to the quantum domain, and Raman and his group took concrete efforts to explore it experimentally. They were aware of the developments due to Einstein, S.N. Bose, Millikan, and Compton, which further motivated him to interrogate the concept of the spin of photons from a light scattering perspective (also see a good overview of physics during Raman's era by Rai Choudhary [31,32] that collectively discusses the science of Raman, Bose, and Saha.

Light Scattering in Amorphous Solids

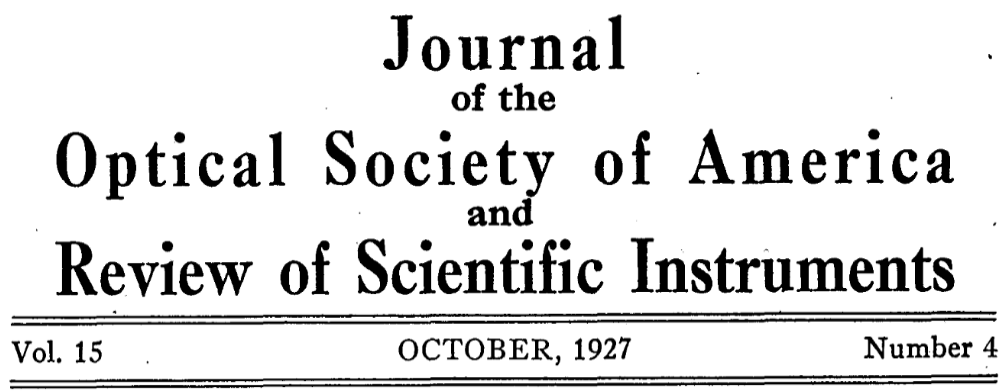

Figure 3. Snapshot of the title of the paper published by Raman in this journal [33].

In 1927, Raman wrote a research paper in the Journal of the Optical Society of America [33], which was then co-published with Review of Scientific Instruments (see figure 3). In there, he discussed the scattering of light in amorphous solids. He introduces the topic with a discussion related to the random orientation of the molecules in a dust-free liquid and then compares it to a mixture of liquids and how both these categories scatter light. When the liquids are cooled down, the change in compressibility and the related modification in the fluctuation density affect light scattering. Such cooled liquid systems that solidify are considered as amorphous solids. Raman was motivated to study what happens to the light scattering in such systems. The questions of intensity, polarization, and wavelength were of interest to him. As he mentions in the introduction of the paper that "we may anticipate that an amorphous solid such as glass, consisting of a mixture of anisotropic molecules, would exhibit when light traverses it, a partially-polarized internal scattering or opalescence of an order of intensity not greatly inferior to that ordinarily observed in liquids or liquid mixtures." (p. 185 in [15])

One of the points he is making here is to emphasize that an amorphous solid would show light scattering properties similar to conventional liquids or liquid mixtures. The role of the molecules and their orientation is an equally important question in this problem. During that era, the contribution of the molecules towards light scattering signatures was yet to be established on a strong footing. Raman was trying to probe this problem and connect them to material composition and their optical properties in a systematic way. As he mentions: "If the phenomenon has a true molecular origin, we should expect to find the intensity of scattering to be definitely correlated with the refractivity and chemical constitution of the glass." (p. 186 in [15])

The paper reported a light scattering study in a series of 14 optical glasses manufactured in Jena, Germany and includes compositions such as Borosilicate crown, ordinary flint, dense flint, prism crown, etc. This systematic study clearly showed that the intensity of the scattered light had a connection to the refractive index and the composition of the material. Raman was interested to know the origin of the scattered light, and he claimed that such a connection would not be possible due to imperfections in the glasses, but rather due to the molecular structure of the material. And he indicates: "the fact that the intensity of scattering is very clearly a function of the refractive-index and chemical composition of the glass, render it extremely improbable that the effect can arise from accidental inclusions or imperfections in the structure of the glasses. It is, in fact, clear from the data that the effect arises from the ultimate molecular structure of glass." ( p. 188 in [15])

## Oblique Incidence Reappears with Metallic Screens

In 1927, Raman and Krishnan published a paper [34] titled "Diffraction of Light by Metallic Screens". The motivation was to explain some of the experimental observations reported by Gouy in 1886. Gouy's experiment was related to the diffraction of light from metallic screens and wedges. He reported diffraction, which had dependencies on the intensity, polarization, and the color of the light. Subsequently, the theoretical analysis was proposed by Poincaré and further analyzed by Sommerfeld. Although these theoretical analyses could explain some of the observations, they were not complete. Raman and Krishnan modified the formula proposed by Poincaré and Sommerfeld by incorporating the contribution of phase and amplitude of the light in the diffraction problem. One of the configurations they discussed was the elliptically polarized light that emerged from the diffraction of plane-polarized light on such metallic screens and wedges. The sign of the ellipticity depended on the internal or the external diffraction configuration. Furthermore, the sign of the ellipticity had dependencies on the nature of the metal considered in the analysis. They also studied imperfect conductors and showed a relationship between oblique incidence angle and diffraction intensity, and connected it to the reflection coefficient of metal, thus showing the relevance of considering material characteristics in diffraction problems [34]. From a modern optics perspective, metallic material and their interaction with light have turned out to be important in the context of nanophotonics, particularly nanoplasmonics [35]. These experiments have interesting connections to such phenomena, and Raman and Krishnan utilized diffraction theory to analyze such effects.

### Bringing Light and Sound together - Acousto-optics (1930 – 1940)

As described previously, Raman was motivated to study all kinds of wave phenomena. This included not only light but also sound waves. In this context, Raman was looking at the interaction of light with sound waves in a medium. In the 1930s, this phenomenon got an impetus thanks to some wonderful experiments in which quartz crystals were utilized to create the piezoelectric effect [36,37]. These crystals were excited using an external electrical source, which further created sound waves in the medium (generally water). Given that Raman had an interest and background in studying sound waves, such an effect caught his attention of Raman. The motivation for his studies was to not only create sound waves in a medium, but also to look at the consequences of light that is propagating through such a medium.

In this context, the question of diffraction of light due to the gratings formed by the sound waves was an interesting topic to explore. As one may observe, this is in anticipation of acousto-optical phenomena, which are now at the heart of a variety of photonic and ultrasonic technologies. Raman, in collaboration with his student Nagedra Nath, had an important role to play in laying the foundations of the acousto-optic effect. One of the earliest questions they were interested in was to know how light gets deviated due to the sound waves in a medium, and how to experimentally observe such kind of an interaction? Motivated by such questions, Raman and Nagendra Nath wrote a series of papers [38–43] on the interaction of light with sound waves, specifically ultrasonic sound waves. In their first of many papers in a series, they elegantly describe the interaction from an experimental viewpoint as follows [38] :

> "The arrangement may be described briefly as follows. A plane beam of monochromatic light emerging from a distant slit and a collimating lens is incident normally on a cell of rectangular cross-section and after passing through the medium, emerges from the opposite side. Under these conditions, the incident beam will be undeviated if the medium be homogeneous and isotropic. If, however, the medium be traversed by high-frequency sound-waves generated, by introducing a quartz oscillator at the top of the cell, the medium becomes stratified into parallel layers of varying refractive index. Considering the case in which the incident beam is parallel to the plane of the sound-waves, the emerging light from the medium will now consist of various beams travelling in different directions."

In 1930, this type of observation was done by various researchers across the world, including the USA and France [38]. Raman and Nath were interested in understanding the physics behind the phenomenon and were trying to come up with a theoretical explanation. Specifically, they were building upon some theories proposed by Debye, Sears, Brillouin, Lucas and Biquard. They mention some of the shortcomings of the proposed theories and were intrigued by the question "why there should be so many orders and why the intensity should wander between the various orders?" (p. 407 in [38])

Motivated by these questions, Raman and Nath proposed a theory that would explain the transmission of light and accompanying phase changes due to high-frequency sound waves in the medium. They go about matching their theoretical formula with experimental observation of other researchers (particularly that of Bär, Helv. Phys. Acta, 1988, 6, 570.) [38]

They conclude their first paper with the following remarks [38]:

"(a) A theory of the phenomenon of the diffraction of light by sound waves of high frequency in a medium, discovered by Debye and Sears and Lucas and Biquard, is developed.

(b) The formula

$$\sin\theta = \pm\frac{n\lambda}{\lambda^*} \qquad n \text{ (an integer)} \geqslant 0$$

which gives the directions of the diffracted beams from the direction of the incident beam and where $\lambda$ and $\lambda^*$ are the wave-lengths of the incident light and the sound wave in the medium, is established. It has been found that the relative intensity of the $m$th component to the $n$th component is given by

$$J_m^{\,2}(2\pi\mu\, L/\lambda)/\, J_n^{\,2}(2\pi\mu\, L/\lambda)$$

where the functions are the Bessel functions of the $m$th order and the $n$th order, $\mu$ is the maximum variation of the refractive index and L is the path traversed by light. These theoretical results interpret the experimental results of Bär in a very gratifying manner."

Raman and Nagendra Nath wrote a series of five papers [38–43] mainly discussing the theoretical model of diffraction of light by sound. This was mainly based on the phase grating model and the resultant effects. Nagendra Nath, who was an excellent theoretician himself, wrote two independent papers as a follow-up to the postulated theory [44,45]. There were also some experimental papers that were published subsequently and were connected to the work of Raman and Nath. All these papers emanated from Raman's group at the Indian Institute of Science.

## Physics of Crystals (1940 -1950)

Raman had a long-standing interest in the physics of crystals [5]. Particularly, he was interested in understanding the interaction of light with gemstones and rare crystals. This was motivated by his already well-founded work on light scattering and his intuition on utilizing the ideas from the molecular light scattering to solid-state materials. When Raman was in Bangalore, his research interest deepened in crystal dynamics. Among the many crystals that he studied, he took special interest in diamond [46]. Raman and his laboratory recorded the famous 1332 cm$^{-1}$ line of the diamond, and the quality of the spectral features was on par with or superior to many of the existing methods in the world. Naturally, Raman and his team were interested in explaining the origins of the spectral features of diamond and were trying to build a model to explain the rich spectral features. In order to understand the spectral features of crystals, one had to further the understanding of crystal dynamics, which was still in a nascent stage in the 1920s and 30s.

Crystal dynamics and thermodynamics have an interesting connection. One of the first aspects of this connection can be found in the Dulong-Petit law, which was postulated in the 1810s. It revealed that the molar heat capacity (specific heat) of many solids had a constant value. This motivated Boltzmann in the 1870s to give a theory from an atomistic viewpoint of vibrations in solids. The theory matched well for certain solids and disagreed with a few elements, including carbon, boron, and silicon(see p. 401 in [9]). Importantly, the specific heat of solids also depended on the temperature, which was not yet explained. Later, around 1907, Einstein utilized Planck's law and discussed the specific heats of solids and described the temperature dependence, although the model used was not realistic. This was followed by Debye's theory (see discussion on p. 401 in reference [9]), which was strongly influenced by Rayleigh's model of a one-dimensional chain of atoms and the related vibration waves of the collection of atoms. In 1912, around the same time as Debye's theory, Born and Von Karman also studied the vibration of the crystal lattice. They used the picture of traveling waves. Importantly, they used the so-called 'periodic or cyclic boundary conditions' [47,48] to compute the frequencies of the vibration in a crystal lattice. It is this periodic boundary condition that came under attack by Raman. Raman had the viewpoint of something called a supercell (for a diagrammatic explanation, see chapter 10 in [9]), and he utilized that to compute the frequencies in the spectral features of diamond. His model had a specific calculation of normal modes and resulted in certain distributions of the peaks. The first approximation matched with the experiments, but with closer scrutiny, one could not account for the broad background in spectral features.

As time progressed, Born's theory and the periodic boundary condition found extensive acceptance. It became an important feature of solid-state physics. Raman's supercell idea was isolated and did not get traction either from theoretical physicists or from solid-state experimentalists. Eventually, Born's theory was found to be correct, and Raman's interpretation had certain flaws. Although Born and Raman had a very good working relationship (In the 1930s, Born and his family spent a few months in Bangalore, and Raman was his host [9]), this difference of opinion on a technical matter affected their communication. Their disagreement [47,48] was also fought out on the pages of various journals, including *Nature* and *Reviews of Modern Physics*.

During this decade, another important contribution that emerged from Raman's group was the study of soft modes [49]. Temperature-dependent spectroscopic investigations on quartz resulted in identifying these low-frequency modes [50]. They have a deep connection to phase transformations, and a variety of crystals and the temperature-dependent spectral features were studied in this context (see p. 341 in reference [9]).

## Speckles in the 1940s

In the 1940s, Raman, along with his student Ramachandran, anticipated the speckle phenomenon much before it became a well-studied concept [51,12]. In the year 1943, Ramachandran from Raman's group at IISc, reported a series of papers [51–53] related to optical fluctuation arising from disordered structures. Figure 4a shows the speckle pattern experimentally observed by Ramachandra, along with the verified Rayleigh's statistical law of fluctuation (see figure. 4b).

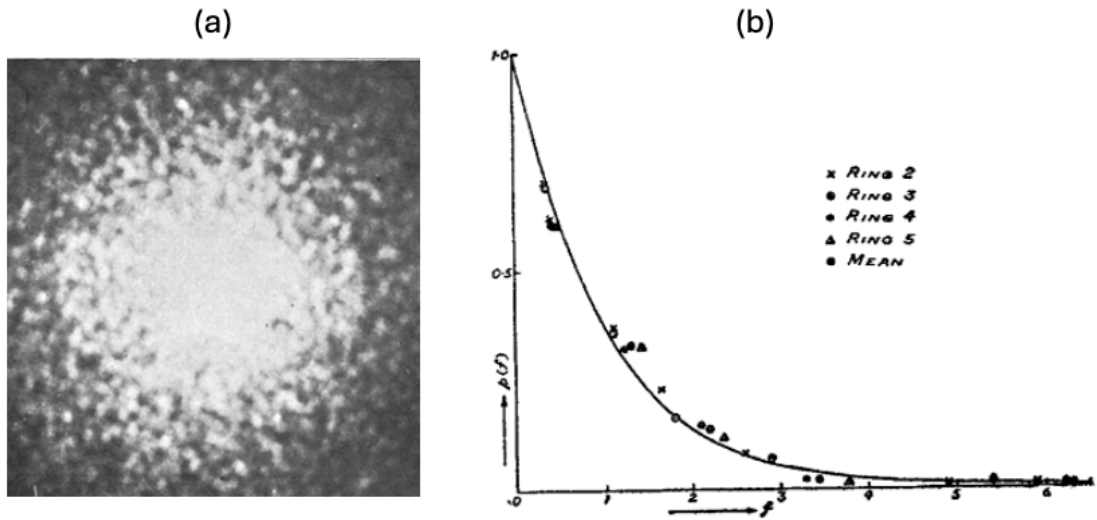

Figure 4. (a) The speckle pattern observed by Ramachandran in the year 1943. The image shows the intensity distribution from the light scattered by a glass plate sprinkled with lycopodium powder. (b) Graph showing the verification of the Rayleigh law of fluctuation. Figures reproduced from the reference [51] under Creative Commons license.

Raman gave a series of lectures [10] on physical optics in which all these concepts, including observations by Ramachandran, were discussed in detail. Writing in 1989, Berry mentions [54] how speckle phenomena were pre-empted by Raman and Ramachandran by utilizing light sources such as a mercury lamp, along with simple optical arrangements:

"I was surprised to learn that he discovered or anticipated several phenomena usually regarded as having been found decades later by others. One of these is what is now called 'speckle'. This is the mottled appearance of coherent light scattered by static randomness, such as grains of powder on a screen or irregularities on a rough painted wall. Nowadays, speckle is a familiar accompaniment of images produced by lasers, but Raman (with Ramachandran) saw it in light from a mercury lamp filtered by a pinhole. He had a complete understanding of the phenomenon, 20 years before lasers were invented." [54]

## Optics of Minerals (1950 – 1960)

Combining aesthetics with optics was one of the common themes of Raman's research. Minerals are a great source for such a combination. In 1948, Raman retired from the Indian Institute of Science and moved to the Raman Research Institute. One of the research areas he pursued during that time was the optics of minerals, as Raman and Rao describe: "In the new Institute he built for himself to work in peace after he formally retired, he started by arranging his magnificent, now famous, collection of minerals, specimens he had gathered from all over the world for the extraordinary optical phenomena they exhibited limestones, marbles, alabaster, gypsums, tourmalines, agates, quartzites, jades, amethysts, fluorites, micas and serpentines, iolites, malachites, lapis lazuli and feldspars." (see page 166 in [10])

The questions of interest during the time were related to optical heterogeneities in minerals, which can be due to variation in composition and refractive index. The passage of light and the subsequent diffusion can facilitate rich information on such heterogeneous media as "they could be isotropic or birefringent, oriented randomly or with a preferred direction; they could show periodicities in one, two or three dimensions. The medium itself could be amorphous or crystalline, polycrystalline or a monocrystal, isotropic or birefringent." [10] Raman and his group were interested in all these aspects and their connection to optics.

## Colour and Human Vision (1960-1970)

Towards the later part of his life, Raman became interested in the perception of colour [6], particularly in the context of the physiology of the eye. Of course, colour in nature was something he was always interested in, but he took a concrete step to explore the physiology of vision from a viewpoint of its optical properties. This led to a series of papers related to this topic and eventually culminated in a monograph titled "The Physiology of Vision" [55].

He explored the optical and spectroscopic properties of visual pigments and those of flowers. In fact, he extensively studied the spectral composition of floral colours. He tried to make a connection to the perception of vision from this information. Many natural objects, including leaves and gemstones, were studied from the optics and spectroscopy viewpoint. Of particular interest to Raman was to understand the functioning of the retina of an eye [48]. He wanted to understand how the eye recognizes colour, brightness, and, under certain conditions, even the polarization of light. Therefore, he took creative methods to probe the optics of the retina with experiments that he himself could perform. In the collected scientific papers of C.V. Raman, Ramaseshan (a scientist and nephew of Raman) introduces Raman's method to probe the retina in the following way [10]:

"Raman realised that if one were to understand human vision, one must explore the retina. He could not put a probe into the eye so he devised a simple method by which one could actually view one's own retina. The observer views a brilliantly illuminated screen, holding before his eye a colour filter (which absorbs completely a limited region of the spectrum while transmitting the rest of it). When the filter is suddenly removed, he sees a highly enlarged view of his own retina (that too in colour) projected on the screen, displaying the response of different areas of it to the incident light. By using a series of filters transmitting different wavelengths, Raman could explore the behaviour of his retina under various spectral excitations." (see p. 174 in [10] and introduction of the reference [6] )

Raman utilized all the methods and techniques he had devised over the years to explore the rich tapestry of colors in natural objects and try to connect them to the perception of vision in a human eye. Of course, many of the theories he was postulating are no longer in use, but some of the questions he raised to understand the spectral characteristics of natural objects remain relevant not only from an optics viewpoint but also in the context of botany and chemical ecology.

Towards the later part of the 1960s, Raman did not have a large research group, but still, he explored a variety of problems motivated by natural optical phenomena. One of the few people whom he mentored during this time was Pancharatnam. In the 1960s, Pancharatnam worked in Raman's lab. They jointly wrote a paper on the optics of mirages [56] . Pancharatnam was a young man in his 20s and was interested in the optics of anisotropic crystals, and he went on to do some outstanding work in this direction [57]. It was in Raman's lab that he undertook the exploration of the optical phase that is now famously known as the Pancharatnam-Berry phase [58–60].

## Conclusion

Apart from being an outstanding scientist, Raman communicated science effectively, both to the specialists and to the public [7]. One of the consistent themes throughout his life was to communicate the excitement of science to the general audience. Raman derived a lot of pleasure in interacting with students and school children, and towards the latter part of his life, he did this frequently (see later chapters of [9] and [10]). Raman also derived a lot of inspiration from the history of science and mentioned this explicitly in many of his talks and lectures. The spirit of science and its connection to the historical overview (of a subject or a scientist) was a common theme in such discussions, and as Raman mentions, [8]"A study of the history of individual branches of science and of the biographies of the leading contributors to their development is essential for a proper appreciation of the real meaning and spirit of science. They often afford much more stimulating reading than the most learned of formal treatises on science. To the teacher, such histories and biographies are invaluable."

This article is my small attempt to uphold Raman's thought, where I have tried to give a glimpse of the history of certain aspects of optical science that Raman was involved in. He and his research group indeed contributed to the development of the field, and it is important for us to appreciate this. In doing so, the hope is to motivate and invite young (and old) people to the vast and beautiful arena of optics and photonics.

Raman's work in optics was an ode to curiosity, which eventually benefited humanity in the form of Raman spectroscopy and its application. His research also showed the deep connection between curiosity-driven research in basic science and societal impact. There is an important lesson for all of us in this.

### Acknowledgements

The author thanks Prof. Anurag Sharma, IIT, Delhi, for the invitation to write this article. He also thanks the Raman Research Institute's online digital depository of archival material on C.V. Raman.